\documentclass[aip,apl,preprint,showpacs]{revtex4}

\usepackage{graphicx}
\usepackage{amssymb}
\usepackage{color}
\usepackage{verbatim}

\begin{document}

\title{Enhancing the stability of a continuous-wave terahertz system by photocurrent normalization}

\author{A. Roggenbuck$^{1,2}$, M. Langenbach$^1$, K. Thirunavukkuarasu$^1$, H. Schmitz$^1$, A. Deninger$^2$,
I. C\'{a}mara Mayorga$^3$, R. G\"{u}sten$^3$,
J. Hemberger$^1$, and M. Gr\"{u}ninger$^1$}

\affiliation{$^1$ II. Physikalisches Institut, Universit\"{a}t zu K\"{o}ln, Z\"{u}lpicher Str.\ 77, D-50937 K\"{o}ln, Germany}
\affiliation{$^2$ TOPTICA Photonics AG, Lochhamer Schlag 19, D-82166 Gr\"{a}felfing, Germany}
\affiliation{$^3$ Max-Planck-Institute for Radio Astronomy, Auf dem H\"{u}gel 69, D-53121 Bonn, Germany}

\begin{abstract}
In a continuous-wave terahertz system based on photomixing, the measured amplitude of the terahertz
signal shows an uncertainty due to drifts of the responsivities of the photomixers and of the optical power
illuminating the photomixers.
We report on a simple method to substantially reduce this uncertainty.
By normalizing the amplitude to the DC photocurrents in both the transmitter and receiver photomixers,
we achieve a significant increase of the stability.
If, e.g., the optical power of one laser is reduced by 10\%, the normalized signal is expected to change
by only 0.3\%, i.e., less than the typical uncertainty due to short-term fluctuations.
This stabilization can be particularly valuable for terahertz applications in non-ideal environmental
conditions outside of a temperature-stabilized laboratory.
% In order to demonstrate the performance of this technique, we purposely induce an uncertainty of
% about 15\% by modulating the laboratory temperature.
% The normalization reduces the uncertainty to a few percent.
\end{abstract}

% OSA Ocis Code
% 300.6495   Spectroscopy, terahertz (300.0300   Spectroscopy)
% 120.6200   Spectrometers and spectroscopic instrumentation (120.0120   Instrumentation, measurement, and metrology)
% 040.2235   Far infrared or terahertz (040.0040   Detectors)
% 230.0250   Optoelectronics (230.0230   Optical devices)
% 140.3490   Lasers, distributed-feedback (140.0140   Lasers and laser optics)

%\pacs{07.57.-c, 32.30.Bv, 78.00.00, 78.47.-p}
% 07.57.-c Infrared, submillimeter wave, microwave and radiowave instruments and equipment
% 32.30.Bv Radio-frequency, microwave, and infrared spectra
% 78.00.00 Optical properties, condensed-matter spectroscopy and other interactions of radiation and particles with condensed matter
% 78.47.-p Spectroscopy of solid state dynamics
% 87.64.-t Spectroscopic and microscopic techniques in biophysics and medical physics

\date{February 14, 2013}
\maketitle

\section{Introduction}

Terahertz spectroscopy requires a reliable determination of the amplitude and phase of the terahertz radiation.
One established method for the generation and coherent detection of continuous-wave terahertz radiation is photomixing
\cite{McIntosh1995, Verghese1998, Deninger2008, Roggenbuck2010, Matsuura2005, Saeedkia2008, Preu2011}.
Typically, terahertz radiation is generated by illuminating a biased photomixer (the transmitter, Tx) with the
optical beat of two near-infrared lasers,
and coherent detection is achieved by measuring the photocurrent $I_\mathrm{THz}$ in a second photomixer
(the receiver, Rx) which is illuminated by both the optical beat and the terahertz radiation.
The accuracy of the quantity of interest, e.g.\ the transmission $T_\mathrm{sample}$ of a given sample,
depends on the stability of the measured terahertz signal $I_\mathrm{THz}$.
This signal in turn depends on the stabilities of the responsivities $S_\mathrm{Tx}$ and $S_\mathrm{Rx}$ of the
two photomixers and on the stabilities of the optical powers of the two lasers at the two photomixers.
All six of these quantities are sensitive to temperature drifts.
Here, we report on an efficient normalization of the terahertz signal by employing \emph{both} photomixers as powermeters.

\section{Concept}

Our aim is to minimize the uncertainty of $I_\mathrm{THz}$.
In order to motivate an expression describing $I_\mathrm{THz}$, we start from the generation of the terahertz wave at the transmitter.
For a constant bias voltage of $U_\mathrm{bias} \approx 10\,\mathrm{V}$,
the transmitter on the one hand behaves like a photoconductive
resistor, i.e.\ the DC photocurrent at the transmitter, $I_\mathrm{Tx}$, is given by the
bias-dependent responsivity $S_\mathrm{Tx}(U_\mathrm{bias})$
and the total incident optical power $P_\mathrm{Tx}=P_\mathrm{1,Tx}+P_\mathrm{2,Tx}$,
\begin{equation}
    I_\mathrm{Tx}=  S_\mathrm{Tx}(U_\mathrm{bias}) \, \times \, (P_\mathrm{1,Tx}+P_\mathrm{2,Tx}) \, ,
    \label{eqn:ITx}
\end{equation}
where $P_\mathrm{1,Tx}$ and $P_\mathrm{2,Tx}$ denote the optical powers of the two lasers at the transmitter.
On the other hand, the interference between the two laser beams gives rise to a beat which yields a current
oscillating at the difference frequency. Thus the transmitter emits a terahertz wave with power
% \begin{equation}
$P_\mathrm{THz,Tx} \propto S^2_\mathrm{Tx}(U_\mathrm{bias}) \, \times \, P_\mathrm{1,Tx} P_\mathrm{2,Tx}$
\cite{Bjarnason2005}.
%    \label{eqn:PTHz}
% \end{equation}
In transmission geometry, the power at the receiver is given by
$P_\mathrm{THz,Rx} \propto \, T_\mathrm{sample} \, \times \, P_\mathrm{THz,Tx}$.
The photocurrent in the receiver, $I_\mathrm{THz} = I_\mathrm{0,THz}\cos(\Delta\varphi)$,
depends on the phase difference $\Delta \varphi$ between the terahertz wave and the optical beat signal
at the receiver.
The information on amplitude $I_\mathrm{0,THz}$ and phase difference $\Delta \varphi$ can be separated by
phase modulation with a mechanical delay stage or, in our case, a fiber stretcher \cite{Roggenbuck2012}.
We focus on the amplitude \cite{Bjarnason2005}
\begin{equation}
   I_\mathrm{0,THz}
   \, \propto \, P_\mathrm{THz,Rx}^{1/2} \, \times \, \frac{\mathrm{d}S_\mathrm{Rx}}{\mathrm{d}U} \, \times \,
   (P_\mathrm{1,Rx} P_\mathrm{2,Rx})^{1/2} \, ,
   \label{eqn:ITHzVsR}
\end{equation}
where $P_\mathrm{i,Rx}$ denote the optical powers of the two lasers at the receiver, and $S_\mathrm{Rx}$ and
$\mathrm{d}S_\mathrm{Rx}/\mathrm{d}U$ are the responsivity of the receiver and its voltage derivative, respectively.
Here, we have used the linear photocurrent-voltage characteristic of a photomixer (see Fig.\ 1), thus
$S_\mathrm{Rx} \, \propto \, E_\mathrm{THz,Rx} \, \times \, \mathrm{d}S_\mathrm{Rx}/\mathrm{d}U$ with the
electric field amplitude $E_\mathrm{THz,Rx} \, \propto \, P_\mathrm{THz,Rx}^{1/2}$.
Equation (\ref{eqn:ITHzVsR}) assumes perfect spatial overlap of the two lasers which is exactly the case
in our setup where the superimposed light is guided in single-mode fibers.
However, if the photomixers are illuminated by a free beam, spatial overlap might have to be considered.
Altogether this yields an expression for the amplitude of the terahertz signal,
\begin{equation}
   I_\mathrm{0,THz} \, \propto \, T_\mathrm{sample}^{1/2} \, \times \,  S_\mathrm{Tx}(U_\mathrm{bias} ) \,
   \frac{\mathrm{d} S_\mathrm{Rx}}{\mathrm{d} U} (P_\mathrm{1,Tx}P_\mathrm{2,Tx}P_\mathrm{1,Rx}P_\mathrm{2,Rx})^{1/2}
    \label{eqn:ITHzVsEField}
\end{equation}

Apparently, $I_\mathrm{0,THz}$ depends on all of the four optical powers $P_{i,j}$, which in principle may drift independently.
A drift of e.g.\ the optical output power of laser 1 affects $P_\mathrm{1,Tx}$ and $P_\mathrm{1,Rx}$
but not $P_\mathrm{2,Tx}$ or $P_\mathrm{2,Rx}$, whereas an attenuation or mechanical displacement of the beam
in only the transmitter branch will change $P_\mathrm{1,Tx}$ and $P_\mathrm{2,Tx}$.
Polarization effects can influence all components differently.
The polarization plays a role here because of the finger structure of our photomixers which more strongly
reflects the component of the $E$ field parallel to the fingers than the perpendicular one (see e.g.\ \cite{Mayorga2007}).

In order to fully compensate the above mentioned drifts, one would have to measure all four optical power components
as well as the responsivities (or $\mathrm{d}S/\mathrm{d}U$),
which is not easily practicable. As a sophisticated alternative, we propose to determine
DC photocurrents in \textit{both} photomixers, treating both of them like photodiodes.
We will show that this enables us to compensate for drifts of the responsivities and of the total optical
powers illuminating the two photomixers.
As a result, the normalized signal is only sensitive to a drift of the relative power of the two
lasers.

\begin{figure}[tb]
\center
  \includegraphics[width=0.7\columnwidth]{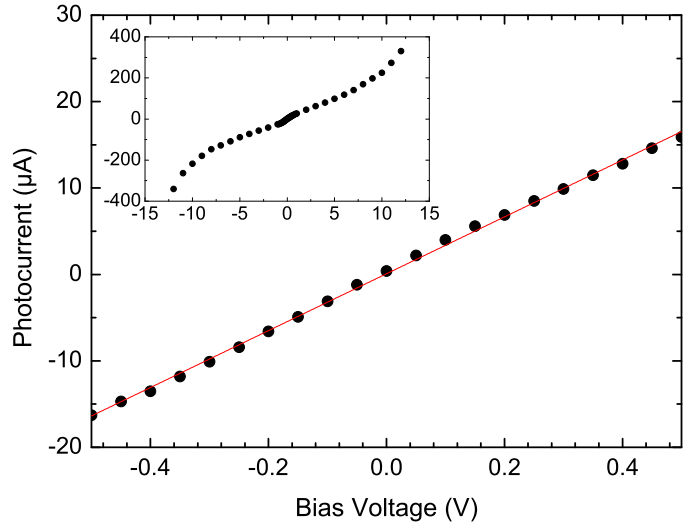}
  \caption{Current-voltage characteristic of an illuminated photomixer.
  At low voltages, the photocurrent is proportional to the voltage (black points: data, red line: linear fit).
  Consequently, the photomixer converts a terahertz electric field linearly into a photocurrent
  (see Eq.\ (\ref{eqn:ITHzVsEField})) and can be used as a detector.
  Inset: Photocurrent-voltage characteristic for larger voltages (see also \cite{Brown1994}).
  Fig1PhotomixerCurrentVoltage.eps.
  }
  \label{fig:IVCharacteristic}
\end{figure}

To the best of our knowledge, biasing a photomixer at the receiver side has not been
reported \cite{Nagatsuma2010}. The transmitter and the unbiased receiver have been discussed above,
see Eqs.\ (\ref{eqn:ITx}) and (\ref{eqn:ITHzVsR}).
Here, we additionally bias the receiver photomixer with a small DC voltage $U_\mathrm{bias,Rx}$ of
typically $\gtrsim$\,30\,mV.\@
Such a small bias does not change the linear behavior of the receiver photomixer (see Fig.\ \ref{fig:IVCharacteristic}),
thus the slightly biased photomixer can still be used as a detector.
For small values of $U_\mathrm{bias,Rx}$ we may write the responsivity as
$S_\mathrm{Rx}(U_\mathrm{bias,Rx}) = U_\mathrm{bias,Rx} \, \times \, \mathrm{d}S_\mathrm{Rx} / \mathrm{d}U$.
Thus the DC photocurrent $I_\mathrm{Rx}$ is given by
\begin{equation}
  I_\mathrm{Rx}= U_\mathrm{bias,Rx}\, \frac{\mathrm{d} S_\mathrm{Rx}}{\mathrm{d} U}\, (P_\mathrm{1,Rx}+P_\mathrm{2,Rx}) \, .
    \label{eqn:IRx}
\end{equation}

Typically, $I_\mathrm{THz}$ (and $I_\mathrm{Tx}$) is not a true DC current because the bias voltage
at the transmitter is modulated e.g.\ in the kHz range in order to apply lock-in detection of $I_\mathrm{THz}$.
Therefore, the DC photocurrent $I_\mathrm{Rx}$ can be separated from $I_\mathrm{THz}$.

Even an unbiased photomixer shows a small residual photocurrent due to a photovoltage which arises
from inhomogeneous illumination \cite{Bjarnason2005}.
Experimentally, we estimate this offset voltage to be of the order of a few millivolt.
The offset voltage itself is sensitive to drifts of the optical power
and thus unsuitable for using the photomixer as a reliable powermeter.
This task requires a stable voltage, independent of the optical power.
Hence, the reliability of the normalization is improved if one selects specimen with low offset voltages,
employs low optical power, and applies a DC bias voltage which is significantly larger than the self-induced
offset voltage of the photomixer.

In short, we determine three different currents.
The photocurrent $I_\mathrm{Tx} \cos(\omega_\mathrm{mod}\, t)$ at the transmitter and the
terahertz photocurrent $I_\mathrm{THz} \cos(\omega_\mathrm{mod}\, t)$ at the receiver
both oscillate at the frequency $\omega_\mathrm{mod}$ of the bias modulation of the transmitter,
whereas $I_\mathrm{Rx}$ is a true DC current due to the DC bias voltage at the receiver.
Then, $I_\mathrm{Tx}$ and $I_\mathrm{Rx}$ are used to normalize $I_\mathrm{THz}$,
\begin{equation}
    I_\mathrm{THz,norm} = I_\mathrm{THz}\, \frac{I_\mathrm{Tx,0}}{I_\mathrm{Tx}}\, \frac{I_\mathrm{Rx,0}}{I_\mathrm{Rx}}
    \propto \, T_\mathrm{sample}^{1/2}\,
    \frac{(P_\mathrm{1,Tx}\, P_\mathrm{2,Tx}\, P_\mathrm{1,Rx}\, P_\mathrm{2,Rx})^{1/2}}{(P_\mathrm{1,Tx}+P_\mathrm{2,Tx})(P_\mathrm{1,Rx}+P_\mathrm{2,Rx})}
    \label{eqn:ITHzNorm}
\end{equation}
where $I_\mathrm{Tx,0}$ and $I_\mathrm{Rx,0}$ are constants, e.g.\ the long-term average values of $I_\mathrm{Tx}$
and $I_\mathrm{Rx}$, respectively.
Note that the normalized photocurrent does not depend on the responsivities $S_\mathrm{Tx}$ and $S_\mathrm{Rx}$ anymore.
This is valid under the assumption that $\mathrm{d}S_\mathrm{Rx} / \mathrm{d}U$ is equal for a constant $U_\mathrm{bias,Rx}$
and for applying a terahertz field $E_\mathrm{THz,Rx} \propto P_\mathrm{THz,Rx}^{1/2}$,
which is supported by our results described below.

Let us now examine how this normalization affects the sensitivity to a drift of the optical power.
The partial derivatives of $I_\mathrm{THz}$ and $I_\mathrm{THz,norm}$ with respect to e.g.\ $P_\mathrm{1,Tx}$
and $P_\mathrm{2,Tx}$ are given by
\begin{equation}
    \frac{\partial I_\mathrm{THz}}{\partial P_\mathrm{1,Tx}}=\frac{1}{2}\frac{I_\mathrm{THz}}{P_\mathrm{1,Tx}}
    \qquad\text{and}\qquad
    \frac{\partial I_\mathrm{THz}}{\partial P_\mathrm{2,Tx}}=\frac{1}{2}\frac{I_\mathrm{THz}}{P_\mathrm{2,Tx}}
    \label{eqn:ITHzDeviation}
\end{equation}
\begin{eqnarray}
    \frac{\partial(I_\mathrm{THz,norm})}{\partial P_\mathrm{1,Tx}} = & - & \frac{1}{2}\frac{I_\mathrm{THz,norm}}{P_\mathrm{1,Tx}}\frac{(P_\mathrm{1,Tx}-P_\mathrm{2,Tx})}{P_\mathrm{1,Tx}+P_\mathrm{2,Tx}}
    \nonumber \\*
    \frac{\partial(I_\mathrm{THz,norm})}{\partial P_\mathrm{2,Tx}} = &   & \frac{1}{2}\frac{I_\mathrm{THz,norm}}{P_\mathrm{2,Tx}}\frac{(P_\mathrm{1,Tx}-P_\mathrm{2,Tx})}{P_\mathrm{1,Tx}+P_\mathrm{2,Tx}}
    \label{eqn:ITHzNormDeviation}
\end{eqnarray}
Compared to $I_\mathrm{THz}$, we find that $I_\mathrm{THz,norm}$ is less sensitive to a variation
of $P_\mathrm{1,Tx}$ by a factor of
$$\left| \frac{P_\mathrm{1,Tx}-P_\mathrm{2,Tx}}{P_\mathrm{1,Tx}+P_\mathrm{2,Tx}} \right|\, .$$
In the limit of $P_\mathrm{1,Tx}=P_\mathrm{2,Tx}$, a small drift of $P_\mathrm{1,Tx}$ is fully compensated.
Moreover, $I_\mathrm{THz,norm}$ also stays constant for a common attenuation of both lasers, i.e.\ for power changes
$\mathrm{d}P_\mathrm{1,Tx}/P_\mathrm{1,Tx}=\mathrm{d}P_\mathrm{2,Tx}/P_\mathrm{2,Tx}$.
However, full compensation is not achieved if $P_\mathrm{1,Tx}$ and $P_\mathrm{2,Tx}$ are significantly
different and, at the same time, change differently.
Therefore, equal laser powers are preferable for the normalization.
An analogous calculation yields the same results for the receiver.

It is instructive to rewrite $I_\mathrm{THz,norm}$ in terms of the splitting ratios of the two laser powers at the transmitter
$r_\mathrm{Tx}=P_\mathrm{1,Tx}/(P_\mathrm{1,Tx}+P_\mathrm{2,Tx})$
and at the receiver
$r_\mathrm{Rx}=P_\mathrm{1,Rx}/(P_\mathrm{1,Rx}+P_\mathrm{2,Rx})$,
\begin{equation}
    I_\mathrm{THz,norm}\propto \, T_\mathrm{sample}^{1/2} \, \times \, [r_\mathrm{Tx}(1-r_\mathrm{Tx})]^{1/2} \, [r_\mathrm{Rx}(1-r_\mathrm{Rx})]^{1/2} \, .
    \label{eqn:ITHzNormSplittingRatio}
\end{equation}
The normalized terahertz photocurrent is no longer a function of the total power or of the responsivities of the photomixers,
but it only depends on the splitting ratios in the two branches.
In the simple case of equal splitting ratios for transmitter and receiver, $r_\mathrm{Tx}=r_\mathrm{Rx} \equiv r$,
Eq. (\ref{eqn:ITHzNormSplittingRatio}) reduces to
\begin{equation}
    I_\mathrm{THz,norm}\propto \, T_\mathrm{sample}^{1/2} \, \times \, r(1-r) \, .
    \label{eqn:r}
\end{equation}

\section{Experimental setup}
\label{sec:Setup}

\begin{figure}[tb]
\center
  \includegraphics[width=0.9\columnwidth]{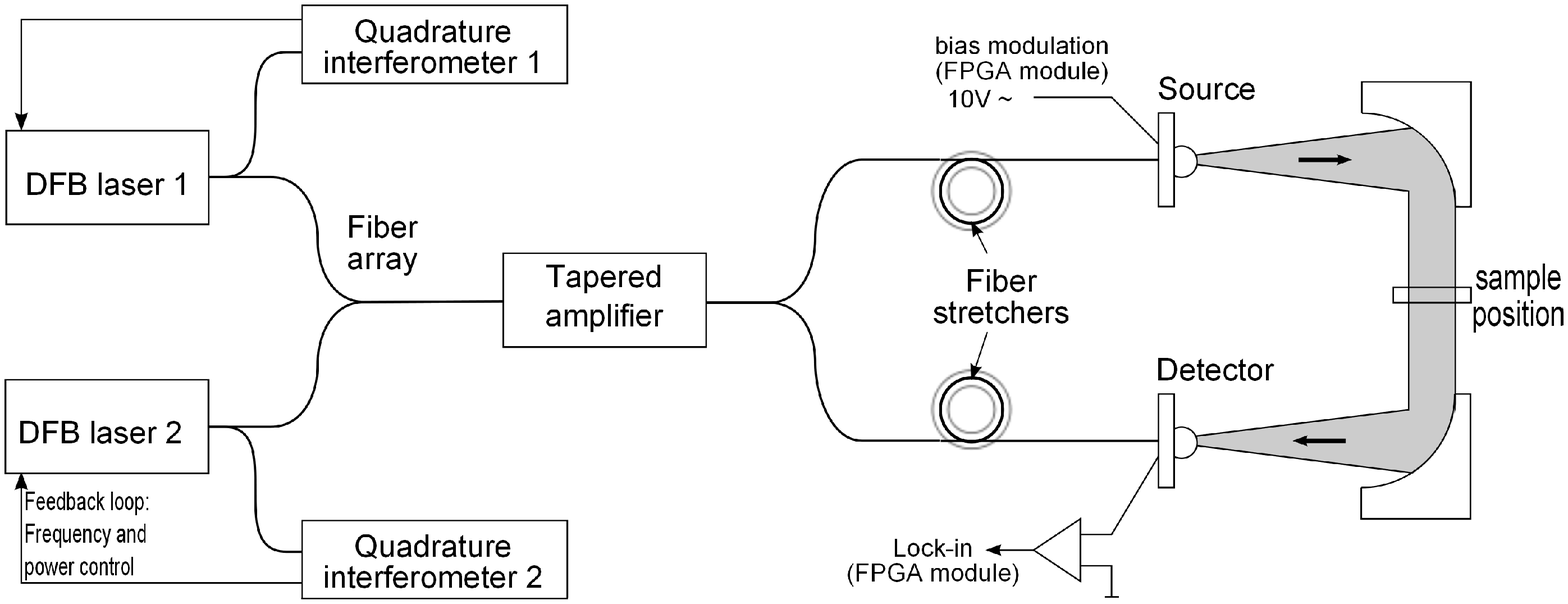}
  \caption{Setup of the continuous-wave terahertz spectrometer (from \cite{Roggenbuck2012}).
  Fig2THzSetupFiberStretcher.eps.
  }
  \label{fig:Setup}
\end{figure}

A sketch of our experimental setup is given in Fig.\ \ref{fig:Setup}, for details we refer to
Refs.\ \cite{Deninger2008,Roggenbuck2010,Roggenbuck2012}.
For both seed lasers, the optical power at a given frequency is actively stabilized to about $\pm$\,1\%,
but the tapered amplifier and the photomixers are sensitive to temperature drifts.
The photocurrent at the receiver is preamplified and then digitized to determine the terahertz photocurrent
$I_\mathrm{THz}$ and the DC receiver photocurrent $I_\mathrm{Rx}$.
The transmitter photocurrent $I_\mathrm{Tx}$ is measured with the help of a 1\,k$\Omega$ resistor in series
with the photomixer.

The signals of both optical frequencies are amplified simultaneously in the tapered amplifier,
and the resulting beam with a particular value of $r$ is sent to both photomixers. Thus the simple case of equal
splitting ratios (see Eq.\ (\ref{eqn:r})) is applicable to our setup. However, small changes of $r_\mathrm{Tx}$
and $r_\mathrm{Rx}$ in the path from the amplifier to the two photomixers cannot be excluded.
The gain of the tapered amplifier is wavelength dependent due to reflections at the surfaces of the amplifier chip.
A temperature-induced change of the chip length therefore leads to a drift of the ratio of the laser
powers which can typically range from 40:60 to 50:50, equivalent to $r = 0.4$ to 0.5.
We thus expect that fluctuations of the normalized photocurrent $I_\mathrm{THz,norm}$ are suppressed to below 5\,\%.

\section{Results}

\begin{figure}[tb]
\center
  \includegraphics[width=0.6\columnwidth]{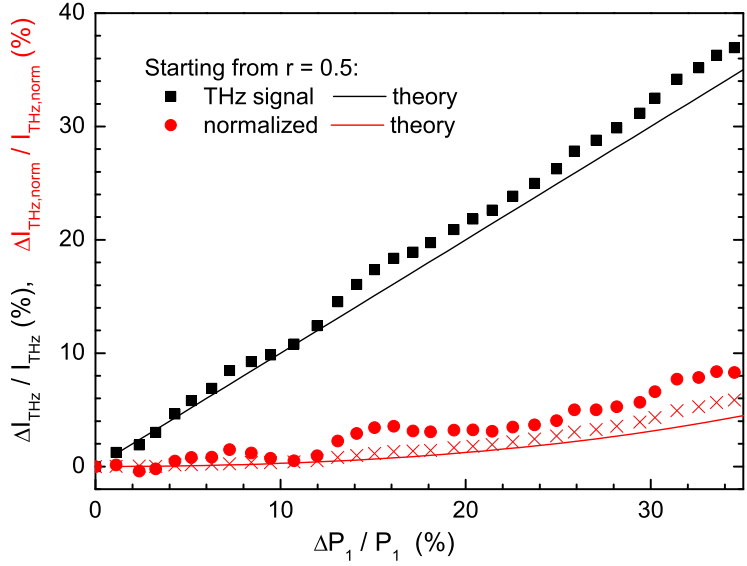}
  \caption{Drift of the measured terahertz photocurrent $I_\mathrm{THz}$ (black) and of the
  normalized photocurrent $I_\mathrm{THz,norm}$ (red) upon reduction of the power $P_1$ of one of the two lasers.
  $\Delta P_1=0$ corresponds to equal laser powers, i.e., $r=0.5$ or $P_1=P_2$.
  Solid lines: predictions according to Eqs.\ (\ref{eqn:ITHzVsEField}) and (\ref{eqn:r}).
  Full symbols: measured data.
  Crosses are obtained by subtracting the deviation between black symbols and black line from the red symbols.
  Fig3normPC.eps.
  }
  \label{fig:normPC}
\end{figure}

As a first example we study the effect of an artificially introduced drift $\Delta P_1$ of the power $P_1$ of
one laser ($P_\mathrm{1,Tx}\propto P_1$ and $P_\mathrm{1,Rx} \propto P_1$).
For this measurement, the tapered amplifier was removed from the setup.
Note that $P_1$ was not measured directly at the photomixers but with a separate photodiode.
Experimentally, we started with equal laser powers, $r=0.5$, and then reduced $P_1$, keeping the power of
the second laser fixed (see Fig.\ \ref{fig:normPC}).
According to Eqs.\ (\ref{eqn:ITHzVsEField}) and (\ref{eqn:r}), we expect that the terahertz photocurrent $I_\mathrm{THz}$
shows the same relative drift as $P_1$, i.e.\ $\Delta I_\mathrm{THz}/I_\mathrm{THz} = \Delta P_1/P_1$,
whereas the relative drift of the normalized current $I_\mathrm{THz,norm}$ is expected to be much smaller (solid lines in
Fig.\ \ref{fig:normPC}).
This is corroborated by the experimental data.
However, the measured drift $\Delta I_\mathrm{THz,norm}$ is slightly larger than expected.
This is mainly due to the fact that $\Delta I_\mathrm{THz}/I_\mathrm{THz}$
shows additional fluctuations of the order of roughly 3\%.
If we subtract these deviations between $\Delta I_\mathrm{THz}/I_\mathrm{THz}$ and the straight line from $\Delta I_\mathrm{THz,norm}/I_\mathrm{THz,norm}$, then the result is very close to the predicted curve (see crosses in Fig.\ \ref{fig:normPC}).
The remaining difference can be attributed to, e.g., the small power-dependent offset voltage at the receiver caused by inhomogeneous illumination.
Although the measured data do not fully reach the ideal case of the theoretical prediction,
the stability of the signal is significantly enhanced by the normalization.
For instance a reduction of $P_1$ of 20\% gives rise to only 3\% change of the measured value of $I_\mathrm{THz,norm}$.

\begin{figure}[tb]
\center
  \includegraphics[width=0.6\columnwidth]{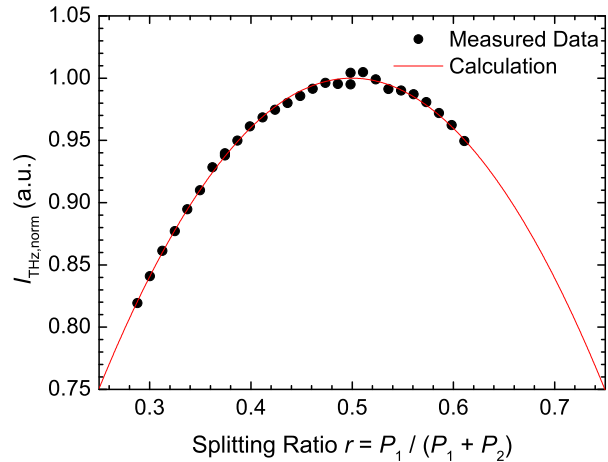}
  \caption{Normalized terahertz photocurrent $I_\mathrm{THz,norm}$ vs. splitting ratio of the two laser powers, $r=P_1/(P_1+P_2)$,
  for a constant total power $P_1+P_2=\text{const}$.
  The solid line depicts our expectation $I_\mathrm{THz,norm} \, \propto \, r(1-r)$, see Eq.\ (\ref{eqn:r}).
  $I_\mathrm{THz,norm}$ is scaled such that $I_\mathrm{THz,norm}(r=0.5)=1$.
  Fig4ITHzVsSplittingRatio.eps.
  }
  \label{fig:SplittingRatio}
\end{figure}

In the second example, see Fig.\ \ref{fig:SplittingRatio}, we changed the splitting ratio $r$
while keeping the total laser power $P_1+P_2$ fixed. In this case, $I_\mathrm{THz,norm}$ follows the predicted
behavior $r(1-r)$ excellently.

\begin{figure}[tb]
\center
  \includegraphics[width=0.6\columnwidth]{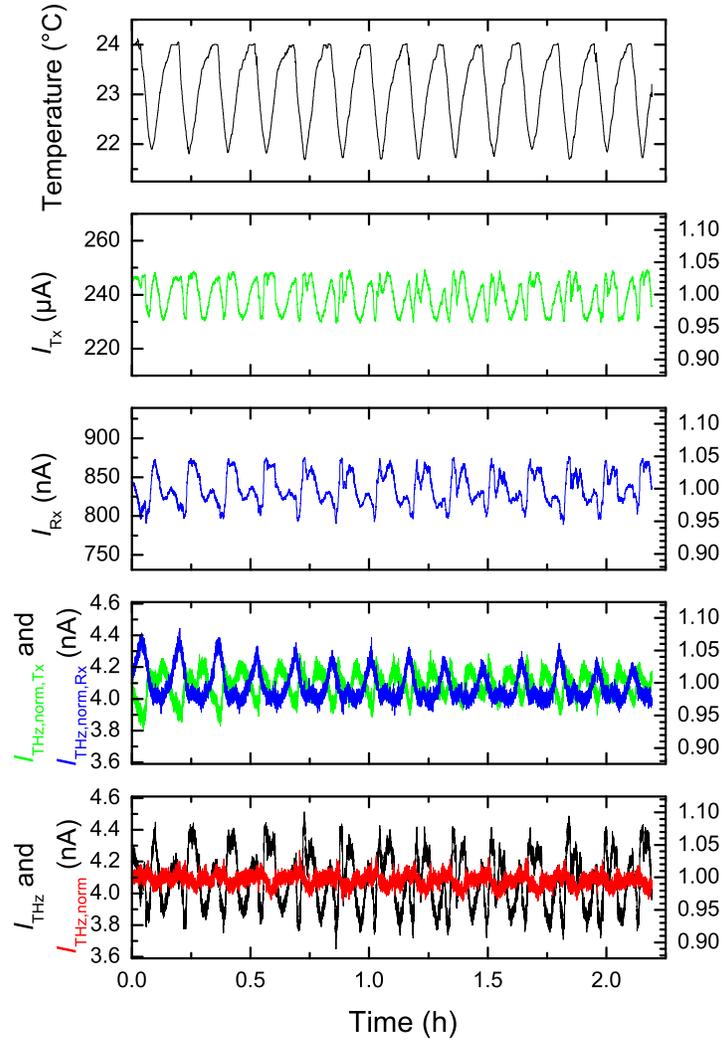}
  \caption{Effect of a periodic variation of the laboratory temperature by $\pm 1$\,K (top panel)
  on the stability of the measured terahertz signal $I_\mathrm{THz}$ (black line, bottom panel).
  The drift is strongly suppressed in the normalized photocurrent $I_\mathrm{THz,norm}$ (red line, bottom panel).
  Second and third panel: DC photocurrents $I_\mathrm{Tx}$ and $I_\mathrm{Rx}$ at transmitter and receiver, respectively.
  Fourth panel: Normalized photocurrent if only $I_\mathrm{Tx}$ (green) or only $I_\mathrm{Rx}$ (blue)
  is used for the normalization,
  e.g.\ $I_\mathrm{THz,norm,Tx} = I_\mathrm{THz} \left(I_\mathrm{Tx,0} / I_\mathrm{Tx}\right)^2$.
  Fig5PhotocurrentCorrectionFigPaper.eps.
  }
  \label{fig:Data}
\end{figure}

As a third example, we deployed the complete spectrometer from Fig.\ \ref{fig:Setup} and monitored
the terahertz amplitude at 600\,GHz over 2 hours while periodically changing the laboratory temperature
between 22\,$^\circ$C and 24\,$^\circ$C (see Fig.\ \ref{fig:Data}).
The measured amplitude varies by approximately 15\% peak-to-peak, whereas the stability of the
normalized amplitude is improved to about 4\% peak-to-peak (bottom panel).
As discussed in Section \ref{sec:Setup}, in our setup the splitting ratio $r$ drifts between 0.4 and 0.5 as a function of temperature.
According to Eq.\ (\ref{eqn:r}), this corresponds to a drift of $I_\mathrm{THz,norm}$ of 4\%, in excellent agreement with our data.
The DC photocurrents in the transmitter and receiver
are correlated to some extent (see Fig.\ \ref{fig:Data}, second and third panel),
which here is mainly due to drifts of the tapered amplifier which is common to both optical paths.
But there are also significant differences in the fluctuations of $I_\mathrm{Tx}$ and $I_\mathrm{Rx}$,
substantiating the necessity to measure both currents separately.
The normalized amplitude is less stable if only $I_\mathrm{Tx}$ or only $I_\mathrm{Rx}$
is used for the normalization. This is evident from the fourth panel of Fig.\ \ref{fig:Data},
which shows $I_\mathrm{THz,norm,Rx}$ and $I_\mathrm{THz,norm,Tx} = I_\mathrm{THz}\,\left(I_\mathrm{Tx,0}/I_\mathrm{Tx}\right)^2$.
In the latter, the factor $I_\mathrm{Rx,0}/I_\mathrm{Rx}$ has been replaced by $I_\mathrm{Tx,0}/I_\mathrm{Tx}$ in Eq. (\ref{eqn:ITHzNorm}).

\begin{figure}[tb]
\center
  \includegraphics[width=0.7\columnwidth]{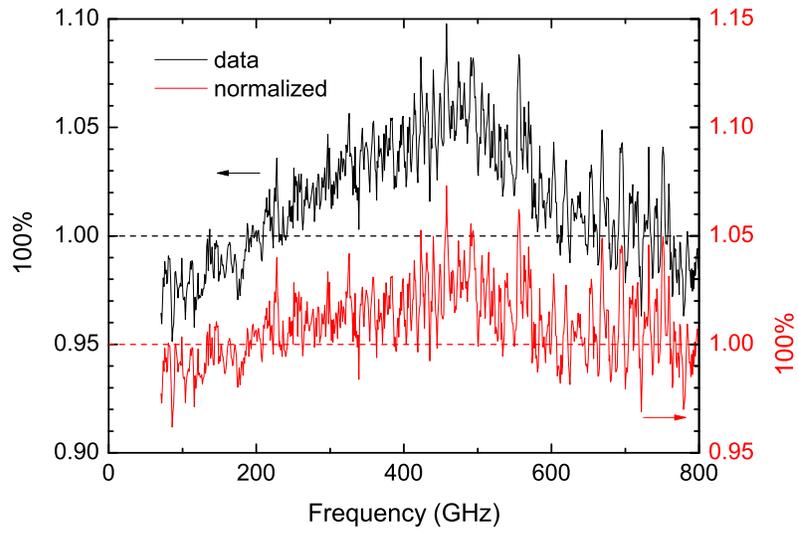}
  \caption{A 100\,\% line, i.e.\ the ratio of the terahertz signals of two consecutive runs,
  measured directly after power up of the system.
  The data have been smoothed by averaging over 5 points.
  Fig6100percent.eps.
  }
  \label{fig:100PercentLine}
\end{figure}

In Fig.\ \ref{fig:100PercentLine} we show the 100\,\% line of the spectrometer in the frequency range from 50\,GHz to 800\,GHz.
The measurements were performed directly after switching on the setup.
For a step size of 1\,GHz, a single run took about 10 minutes,
and there was a delay of about 10 minutes between the two runs.
The initial change of the temperature gives rise to a drift of the 100\,\% line,
which is significantly reduced by the normalization.
As discussed above for the data of Fig.\ \ref{fig:Data}, the normalization is very well suited for temperature-induced drifts.
However, it does not have any effect on the noise-like features present in the 100\,\% line.
These are caused by e.g.\ the uncertainty in the determination of the amplitude $I_\mathrm{0,THz}$ from the raw data measured as a function of $\Delta \varphi$ (see Eq.\ (\ref{eqn:ITHzVsEField})) and by small fluctuations of the frequency.

Finally, we varied the receiver bias and measured the noise photocurrent, i.e.\ the standard deviation of the
terahertz photocurrent with blocked terahertz beam. We found that the noise photocurrent depends on the details
of the photomixer device. For some devices, the noise is nearly independent of the bias, while other devices
show a significant increase of the noise photocurrent when a bias of a few 10\,mV is applied.
Presumably, this difference originates in differences in the photomixer resistivity.

In order to obtain stable and reliable data, it is of course desirable to reduce the fluctuations
of the laboratory temperature in the first place. However, stabilization to much better than $\pm 1\,K$
is not a trivial task.
Moreover, the stabilization achieved via photocurrent normalization may be a significant
improvement for terahertz applications requiring measurements outside a regular laboratory.
As an alternative to the method described here, one may consider to monitor the optical power of the
two lasers with, e.g., a photodiode.
However, the discussed normalization via the photocurrents has two main advantages.
It does not require an extra sensor, and the power is measured directly within the photomixers.
Therefore, the normalization also compensates drifts of the responsivity of the
photomixers, e.g. caused by mechanical drifts within the photomixer device.

\section{Conclusion}

We have described and demonstrated a normalization scheme for the terahertz photocurrent in a continuous-wave photomixing system.
This method is based on measuring the DC photocurrents in both the transmitter and receiver photomixers.
Consequently, no extra sensor is needed and the method can easily be implemented.
Any change of the laser power illuminating the photomixers can be described as a change of the
total laser power in combination with a changing splitting ratio $r$ of the two laser powers.
The normalization fully compensates drifts of the total laser power as well as drifts of the responsivities, thus the stability of the normalized terahertz signal only depends on the splitting ratio $r$.
We have provoked large changes of the terahertz signal by either reducing the power of one laser or by changing the laboratory temperature to simulate unstable ambient conditions.
In all cases, the normalized signal is stable within a few percent.

\section*{Acknowledgements}
This project is supported by the DFG via SFB 608.

\clearpage

\section*{List of Figure Captions}

          Fig.\ 1. Current-voltage characteristic of an illuminated photomixer.
  At low voltages, the photocurrent is proportional to the voltage (black points: data, red line: linear fit).
  Consequently, the photomixer converts a terahertz electric field linearly into a photocurrent
  (see Eq.\ (\ref{eqn:ITHzVsEField})) and can be used as a detector.
  Inset: Photocurrent-voltage characteristic for larger voltages (see also \cite{Brown1994}).

\noindent Fig.\ 2. Setup of the continuous-wave terahertz spectrometer (from \cite{Roggenbuck2012}).

\noindent Fig.\ 3. Drift of the measured terahertz photocurrent $I_\mathrm{THz}$ (black) and of the
  normalized photocurrent $I_\mathrm{THz,norm}$ (red) upon reduction of the power $P_1$ of one of the two lasers.
  $\Delta P_1=0$ corresponds to equal laser powers, i.e., $r=0.5$ or $P_1=P_2$.
  Solid lines: predictions according to Eqs.\ (\ref{eqn:ITHzVsEField}) and (\ref{eqn:r}).
  Full symbols: measured data.
  Crosses are obtained by subtracting the deviation between black symbols and black line from the red symbols.

\noindent Fig.\ 4. Normalized terahertz photocurrent $I_\mathrm{THz,norm}$ vs. splitting ratio of the two laser powers, $r=P_1/(P_1+P_2)$,
  for a constant total power $P_1+P_2=\text{const}$.
  The solid line depicts our expectation $I_\mathrm{THz,norm} \, \propto \, r(1-r)$, see Eq.\ (\ref{eqn:r}).
  $I_\mathrm{THz,norm}$ is scaled such that $I_\mathrm{THz,norm}(r=0.5)=1$.

\noindent Fig.\ 5. Effect of a periodic variation of the laboratory temperature by $\pm 1$\,K (top panel)
  on the stability of the measured terahertz signal $I_\mathrm{THz}$ (black line, bottom panel).
  The drift is strongly suppressed in the normalized photocurrent $I_\mathrm{THz,norm}$ (red line, bottom panel).
  Second and third panel: DC photocurrents $I_\mathrm{Tx}$ and $I_\mathrm{Rx}$ at transmitter and receiver, respectively.
  Fourth panel: Normalized photocurrent if only $I_\mathrm{Tx}$ (green) or only $I_\mathrm{Rx}$ (blue)
  is used for the normalization,
  e.g.\ $I_\mathrm{THz,norm,Tx} = I_\mathrm{THz} \left(I_\mathrm{Tx,0} / I_\mathrm{Tx}\right)^2$.

\noindent Fig.\ 6. A 100\,\% line, i.e.\ the ratio of the terahertz signals of two consecutive runs,
  measured directly after power up of the system.
  The data have been smoothed by averaging over 5 points.

%\listoffigures

\clearpage

\end{document}